# Ultra-Broadband Radial Polarization Conversion based on Goos-Hanchen Shift

*P.B. Phua[1,2] and W.J. Lai[2]*

[1] *DSO National Laboratories, 20 Science Park Drive, Singapore 118230.*
[2] *Nanyang Technological University, 1 Nanyang Walk, Blk 5 Level 3, Singapore 637616.*

**Abstract:**

We demonstrate, for the first time, a scheme that generates radially-polarized light using Goos-Hanchen shift of a cylindrically symmetric Total Internal Reflection. It allows ultra-broadband radial polarization conversion for wavelengths differing >1 micron.

# Ultra-Broadband Radial Polarization Conversion based on Goos-Hanchen Shift

*P.B. Phua[1,2] and W.J. Lai[2]*

[1] DSO National Laboratories, 20 Science Park Drive, Singapore 118230.
[2] Nanyang Technological University, 1 Nanyang Walk, Blk 5 Level 3, Singapore 637616.

Goos-Hanchen shift of Total Internal Reflection (TIR) is different for the s- and p- polarization [1]. Such differential phase shift results in polarization transformation with insignificant wavelength-dependence, and therefore, has been employed in broadband polarization optics such as Fresnel Rhomb.

In this paper, we demonstrate, for the first time, a scheme that uses Goos-Hanchen shift to generate radially polarized light. The main advantages of the proposed scheme are 1) it is ultra-broadband and allows radial polarization conversion even for wavelengths differing by >1 micron, 2) it generates an exact continuous radial polarization profile, 3) it allows high power handling since all optics used can be of high damage thresholds.

The essence of the idea is to have a cylindrically-symmetric surface for TIR. One example of such cylindrically-symmetric surfaces is a cylinder with polished barrel for TIR as shown in Fig 1. It is worth noting that the s- and p-polarization of the TIR of such surface, corresponds to the eigen-axis of the polarization transformation and it also has cylindrical symmetry as shown in Figure 2. The dashed arrows in Figure 2 show the radial p-polarization axis. With an input beam of circular polarization, the angle of reflection is designed so that the total TIR gives an accumulated differential Goos-Hanchen phase shift of $90°$. This results in the output State-Of-Polarization (SOP) as shown as solid arrows in Figure 2. Thus, a $45°$ optical activity quartz rotator can subsequently transform such polarization profile into radial polarization.

In our experiment, we use a BK7 cylinder (10.2 mm diameter and 80 mm length) with a polished barrel. Figure 3 shows the differential Goos-Hanchen phase shift for various angles of reflection in BK 7 for three wavelengths that span from UV to mid-IR spectrum. The slight wavelength dependence of the graphs illustrates its broadband characteristics. We designed our polished cylinder to accommodate three TIRs with an angle of reflection of $69.1°$. This gives an accumulated differential Goos-Hanchen phase shift between the s and p- polarization of $90°$. A BK7 plano-convex axicon of cone angle of $102°$ is used to launch the input circularly polarized 532nm light into the cylinder. Upon leaving the cylinder, the output light is re-collimated using another $102°$ BK7 axicon, and subsequently passes through a $45°$ optical activity quartz rotator. Figure 4 shows the generated radially polarized beam observed with and without a polarizer. Observation of the two-arcs rotating with the transmitting axis of the polarizer, clearly demonstrates the radially polarized beam.

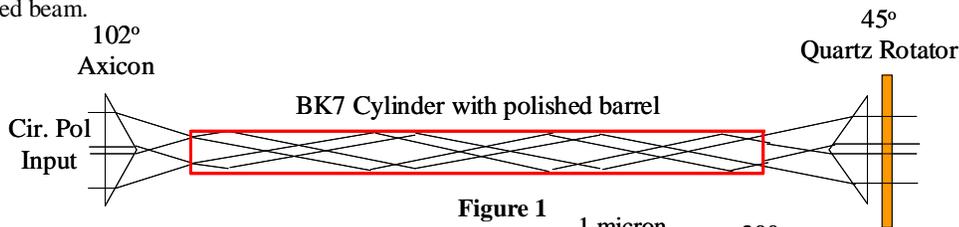

**Figure 1**

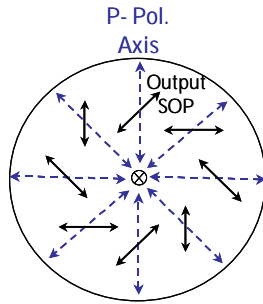

**Figure 2**

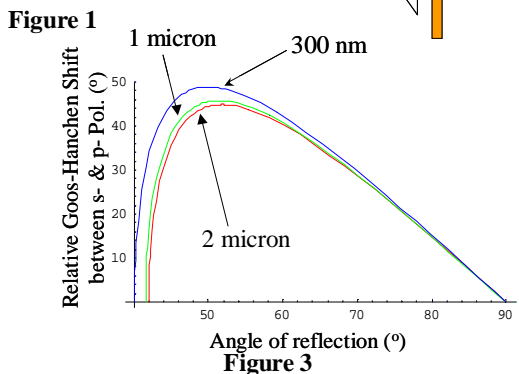

**Figure 3**

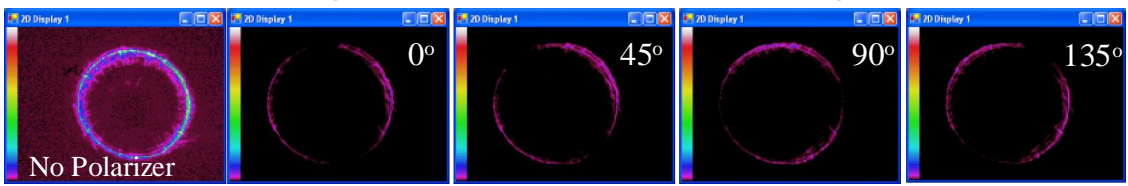

**Figure 4**